\documentclass[aps,prc,twocolumn,superscriptaddress,showpacs,showkeys]{revtex4}
\usepackage{hyperref}

\usepackage{graphicx}
\usepackage{dcolumn}
\usepackage{bm}

\usepackage{float}

\usepackage[usenames]{color}

\begin{document}


\title{Evidence  of Ni break-up from total production cross sections in p+Ni collisions}



\author{A.Budzanowski}
\affiliation{H. Niewodnicza{\'n}ski Institute of Nuclear Physics
PAN, Radzikowskiego 152, 31342 Krak{\'o}w, Poland}
\author{M.Fidelus}
\affiliation{M. Smoluchowski Institute of Physics, Jagellonian
University, Reymonta 4, 30059 Krak{\'o}w, Poland}
\author{D.Filges}
\affiliation{Institut f{\"u}r Kernphysik, Forschungszentrum
J{\"u}lich, D-52425 J{\"u}lich, Germany}
\author{F.Goldenbaum}
\affiliation{Institut f{\"u}r Kernphysik, Forschungszentrum
J{\"u}lich, D-52425 J{\"u}lich, Germany}
\author{H.Hodde}
\affiliation{Institut f{\"ur} Strahlen- und Kernphysik, Bonn
University,  D-53121 Bonn, Germany}
\author{L.Jarczyk}
\affiliation{M. Smoluchowski Institute of Physics, Jagellonian
University, Reymonta 4, 30059 Krak{\'o}w, Poland}
\author{B.Kamys}  \email[Corresponding author: ]{ufkamys@cyf-kr.edu.pl}
\affiliation{M. Smoluchowski Institute of Physics, Jagellonian
University, Reymonta 4, 30059 Krak{\'o}w, Poland}
\author{M.Kistryn}
\affiliation{H. Niewodnicza{\'n}ski Institute of Nuclear Physics
PAN, Radzikowskiego 152, 31342 Krak{\'o}w, Poland}
\author{St.Kistryn}
\affiliation{M. Smoluchowski Institute of Physics, Jagellonian
University, Reymonta 4, 30059 Krak{\'o}w, Poland}
\author{St.Kliczewski}
\affiliation{H. Niewodnicza{\'n}ski Institute of Nuclear Physics
PAN, Radzikowskiego 152, 31342 Krak{\'o}w, Poland}
\author{A.Kowalczyk}
\affiliation{M. Smoluchowski Institute of Physics, Jagellonian
University, Reymonta 4, 30059 Krak{\'o}w, Poland}
\author{E.Kozik}
\affiliation{H. Niewodnicza{\'n}ski Institute of Nuclear Physics
PAN, Radzikowskiego 152, 31342 Krak{\'o}w, Poland}
\author{P.Kulessa}
\affiliation{H. Niewodnicza{\'n}ski Institute of Nuclear Physics
PAN, Radzikowskiego 152, 31342 Krak{\'o}w, Poland}
\affiliation{Institut f{\"u}r Kernphysik, Forschungszentrum
J{\"u}lich, D-52425 J{\"u}lich, Germany}
\author{H.Machner}
\affiliation{Institut f{\"u}r Kernphysik, Forschungszentrum
J{\"u}lich, D-52425 J{\"u}lich, Germany}
\author{A.Magiera}
\affiliation{M. Smoluchowski Institute of Physics, Jagellonian
University, Reymonta 4, 30059 Krak{\'o}w, Poland}
\author{B.Piskor-Ignatowicz}
\affiliation{M. Smoluchowski Institute of Physics, Jagellonian
University, Reymonta 4, 30059 Krak{\'o}w, Poland}
\affiliation{Institut f{\"u}r Kernphysik, Forschungszentrum
J{\"u}lich, D-52425 J{\"u}lich, Germany}
\author{K.Pysz}
\affiliation{H. Niewodnicza{\'n}ski Institute of Nuclear Physics
PAN, Radzikowskiego 152, 31342 Krak{\'o}w, Poland}
\affiliation{Institut f{\"u}r Kernphysik, Forschungszentrum
J{\"u}lich, D-52425 J{\"u}lich, Germany}
\author{Z.Rudy}
\affiliation{M. Smoluchowski Institute of Physics, Jagellonian
University, Reymonta 4, 30059 Krak{\'o}w, Poland}
\author{R.Siudak}
\affiliation{H. Niewodnicza{\'n}ski Institute of Nuclear Physics
PAN, Radzikowskiego 152, 31342 Krak{\'o}w, Poland}
\affiliation{Institut f{\"u}r Kernphysik, Forschungszentrum
J{\"u}lich, D-52425 J{\"u}lich, Germany}
\author{M.Wojciechowski}
\affiliation{M. Smoluchowski Institute of Physics, Jagellonian
University, Reymonta 4, 30059 Krak{\'o}w, Poland}

\collaboration{PISA - \textbf{P}roton \textbf{I}nduced
\textbf{S}p\textbf{A}llation collaboration}

\date{\today}

\begin{abstract}
The total production cross sections of light charged particles
(LCPs), intermediate mass fragments (IMFs) and heavy reaction
products of p+Ni collisions available in the literature have been
compared with predictions of a two-step model in the proton beam
energy range from reaction threshold up to $\sim$~3~ GeV. Model
cross sections were calculated assuming, that the reaction proceeds
via an intranuclear cascade of nucleon-nucleon collisions followed
by evaporation of particles from an equilibrated, heavy target
residuum. The shape of the excitation functions was well described
by model calculations for all reaction products. The magnitude of
the cross sections was reasonably well reproduced for heavy reaction
products, i.e. for nuclei heavier than Al, but the cross sections
for lighter products were \emph{systematically} underestimated. This
fact  was used as an argument in favor of a significant break-up
contribution to the reaction mechanism. The present conclusions are
supported by recently published results of investigations of
differential cross sections in p+Ni collisions, which showed that
hypothesis of the break-up of target nucleus is indispensable for a
good reproduction of $d^2 \sigma/d\Omega dE$ for LCPs and IMFs.
\end{abstract}



\maketitle


\section{\label{sec:introduction} Introduction}

 The aim of the present
paper is to show the arguments, based on a systematic review of the
energy dependence of total production cross sections in p+Ni system,
that a break-up of the target nucleus  is responsible for a large
part of the total production cross sections of nuclides with the
mass number smaller than A $\sim$ 30.   It was shown in recent
studies  of p+Ni collisions at several energies
\cite{BUD09A,BUD09B}, that the break-up assumption leads to a very
good description of energy and angular dependencies of
$d^2\sigma/d\Omega dE$ of light charged particles (LCPs), i.e.,
particles with Z$\leq$2, and intermediate mass fragments (IMFs),
i.e., particles with Z$>$2 but smaller than eventual fission
fragments.  The same effect has been previously observed for p+Au
collisions \cite{BUB07A,BUD08A}. The p+Ni nuclear system has been
chosen for the present analysis, because for this system the total
production cross sections were measured in a broad range of ejectile
masses \cite{AMM08A,GRE88A,MIC95A,MIC84A,SCH96A,RAI75A,REG79A}.
These data together with total production cross sections for LCPs
and light IMFs, determined recently at 0.175, 1.2, 1.9, and 2.5 GeV
\cite{BUD09A,BUD09B} by our group at the COSY accelerator in
Research Center J\"ulich, should allow for a systematic survey
extended to the full range of masses of produced nuclides.

The present paper is organized as follows:   Experimental data are
presented in the next section where they are also compared with the
theoretical cross sections evaluated in the frame of a traditional
two-step model. The discussion of the results is presented also in
this section whereas  the summary with conclusions is given in the
third section.


\section{\label{sec:experiment}Experimental data and their description by a two-step model}

\begin{figure}
\begin{center}
\includegraphics[angle=0,width=0.5\textwidth]{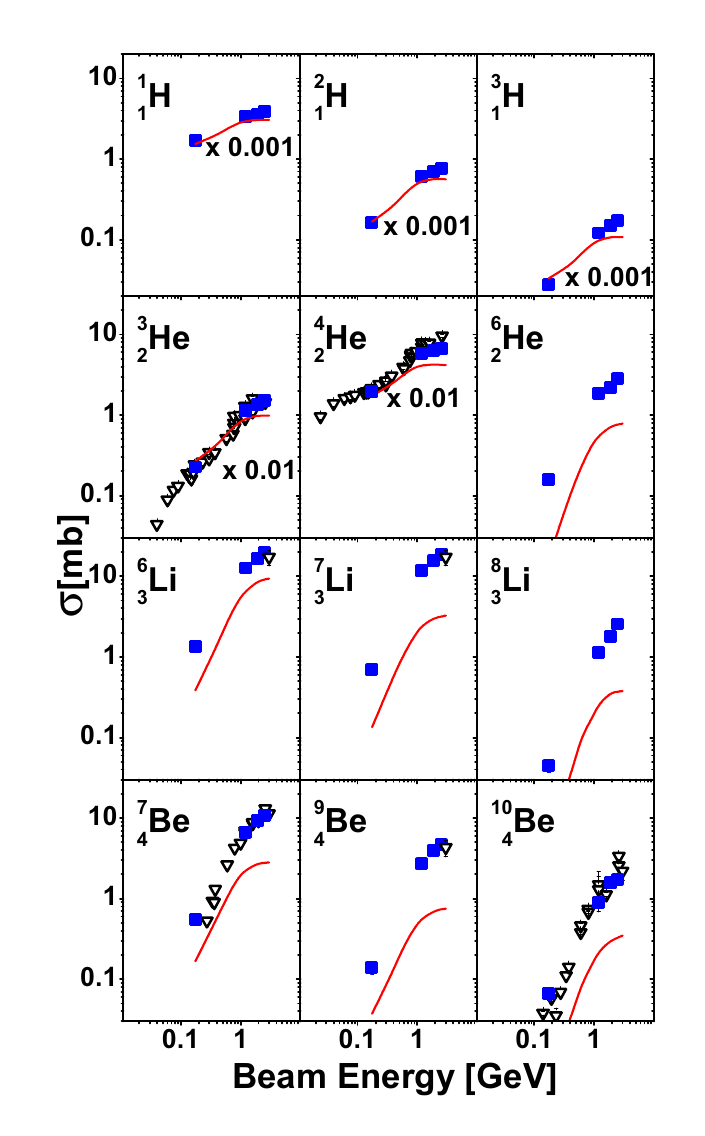}
\caption{\label{fig:h1be10} (Color online) Excitation functions of
production cross sections for $^{1,2,3}H$, $^{3,4,6}$He,
$^{6,7,8}$Li, and $^{7,9,10}$Be in p+Ni collisions. The
symbols represent literature data 
whereas the solid lines depict theoretical excitation functions
which were obtained due to calculations described in the text.  }
\end{center}
\end{figure}

\begin{figure}
\begin{center}
\hspace*{-0.5cm}
\includegraphics[angle=0,width=0.52\textwidth]{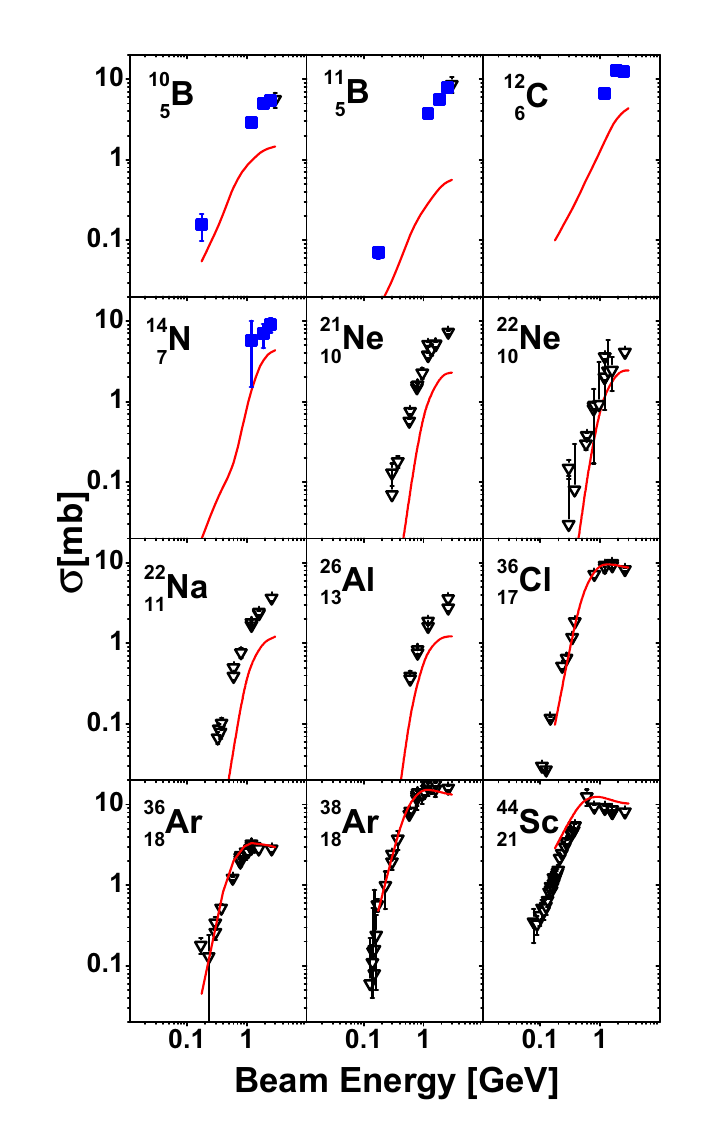}
\caption{\label{fig:b10sc44} (Color online) The same as on Fig.
\ref{fig:h1be10} but for $^{10,11}$B,$^{12}$C,$^{14}$N,
$^{21,22}$Ne,$^{22}$Na, $^{26}$Al, $^{36}$Cl,$^{36,38}$Ar, and
$^{44}$Sc.}
\end{center}
\end{figure}

\begin{figure}
\begin{center}
\hspace*{-0.5cm}
\includegraphics[angle=0,width=0.5\textwidth]{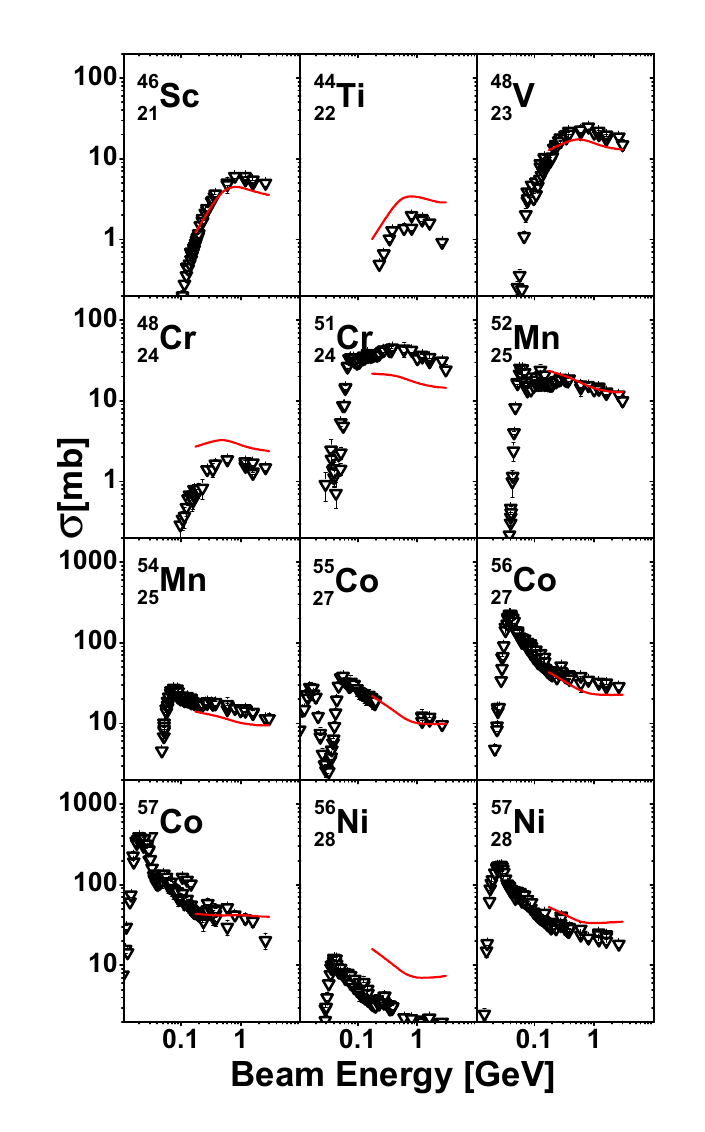}
\caption{\label{fig:scr46ni57} (Color online) The same as on Fig.
\ref{fig:h1be10} but for $^{46}$Sc,  $^{44}$Ti, $^{48}$V,
$^{48,51}$Cr, $^{52,54}$Mn, $^{55,56,57}$Co, and $^{56,57}$Ni. }
\end{center}
\end{figure}

The total cross sections for products of p+Ni collisions at proton
beam energies from the reaction thresholds up to $\sim$ 3 GeV
available in the literature are presented on Figs. \ref{fig:h1be10},
\ref{fig:b10sc44}, and \ref{fig:scr46ni57} for LCPs and light IMFs,
heavier IMFs, and target-like products, respectively.  The data
depicted by open triangles originate from experiments which measured
total cross sections in a straightforward way
\cite{AMM08A,GRE88A,MIC95A,MIC84A,SCH96A,RAI75A,REG79A} whereas the
full squares are the result of the analysis of double differential
cross sections integrated over angle and energy of ejectiles
\cite{BUD09A,BUD09B}. The latter data agree perfectly with former
ones for theses ejectiles where both methods were applied as, e.g.,
for $^{3,4}$He and $^{7,10}$Be. Thus, it can be conjectured, that
they are equally trustworthy for other products. The advantage of
the latter data is that they were measured for such LCPs and light
IMFs, where other methods could not be applied. Due to this, a
representative set of reaction products, covering the full range of
ejectile masses, was collected.

The experimental excitation curves for all products lighter than
$^{44}$Sc increase in the studied energy range, indicating the
beginning of a leveling of the cross sections at the highest
energies (above $\sim$ 1 - 2 GeV).  For heavier products the rise of
the cross section finishes at lower energies, the lower for the
heavier product.  The cross sections of these heavy products
decrease monotonically after reaching a maximum and above 1 GeV a
leveling of the excitation curves starts to appear.

The solid lines shown on the figures for a beam energy higher than
0.175 GeV represent calculations performed in the frame of the
two-step model. The first step -- the intranuclear cascade of
nucleon-nucleon collisions with inclusion of possibility to emit
complex LCPs due to coalescence of nucleons -- was evaluated by
means of the computer program INCL4.3 of Boudard \emph{et al.}
\cite{BOU04A}, and the second step -- evaporation of particles from
an equilibrated residuum of the cascade -- by the GEM2 program of
Furihata \cite{FUR00A,FUR02A}.  The default parameter values,
proposed by the authors of both programs, have been used,
respectively.  The lowest beam energy was chosen to be 0.175 GeV,
because it is the energy at which double differential cross sections
$d^2\sigma/d\Omega dE$ were measured and analyzed using these
programs with  success \cite{BUD09A}.  Still lower energies were not
considered because it is improbable that intranuclear cascade can be
applied at such low energies where the length of de Broglie wave
starts to be too large for treating the nucleus as a set of
individual nucleons - a prerequisite for the applicability of that
model.

The comparison of theoretical lines with the experimental points
leads to the following observations:
\begin{description}
  \item[(i)] The shape of all excitation curves is properly reproduced by
the two-step model.
  \item[(ii)] The magnitude of the cross sections is in
average well reproduced for reaction products  heavier than
A~$\sim$~30 (sometimes one observes even perfect reproduction like,
e.g., for $^{36}$Cl,$^{36,38}$Ar or $^{44}$Sc).
  \item[(iii)] Cross sections for all
LCPs and IMFs lighter than A $\sim$ 30 are \emph{systematically}
underestimated by the model.
\end{description}

The first two of the above observations suggest that the two-step
model properly takes into account a dominating part of the reaction
mechanism for heavy products.  The third observation leads to the
conclusion that for light products a different reaction mechanism
plays the essential role.

It is reasonable to conjecture, that target-like nuclei are produced
as remnants of the evaporation of light particles from an excited
residuum of the intranuclear cascade.  Such a typical spallation
process 
should be well reproduced by the two-step model.  However, the IMFs
may be produced by various processes, i.e., they may be due to
evaporation, may appear as heavy remnants of the evaporation or may
occur from fragmentation of the target. The first two of these
processes are taken into account in the two-step model but the third
mechanism is not considered.  Thus the fragmentation seems to be an
obvious candidate for the process responsible for the observed
inconsistency of the description of total production cross sections.

It was observed (cf., e.g., Ref. \cite{VIO06A}), that for proton
induced reactions at energies of the order of 10 GeV, a
multifragmentation appears with a possible interpretation in terms
of a nuclear liquid –- gas phase transition. However, the enhanced
-- above the predictions of the two-step model -- production of IMFs
and LCPs is visible in Figs. \ref{fig:h1be10} and \ref{fig:b10sc44}
also at much lower energies.  Moreover, the multifragmentation
mentioned above should proceed as the emission from a \emph{single
source} whereas it was shown in recent studies of p+Ni collisions
\cite{BUD09A,BUD09B} that at energies smaller or equal to 2.5 GeV
\emph{two sources} of IMFs are observed. These sources were
interpreted as prefragments of the target appearing due to its
break-up. Since a source can emit only particles with masses smaller
than its own mass, the emission of particles from a single,
target-like source contributes to the yield of all products whereas
the emission from prefragments of the target is limited to products
lighter than the heavier of both prefragments. Thus, only the
break-up hypothesis is compatible with the fact, that a systematic
enhancement of the experimental cross section above the predictions
of two-step model appears for products with masses smaller than
A~$\sim$~30 but not for heavier products.

\begin{figure}
\begin{center}
\includegraphics[angle=0,width=0.50\textwidth]{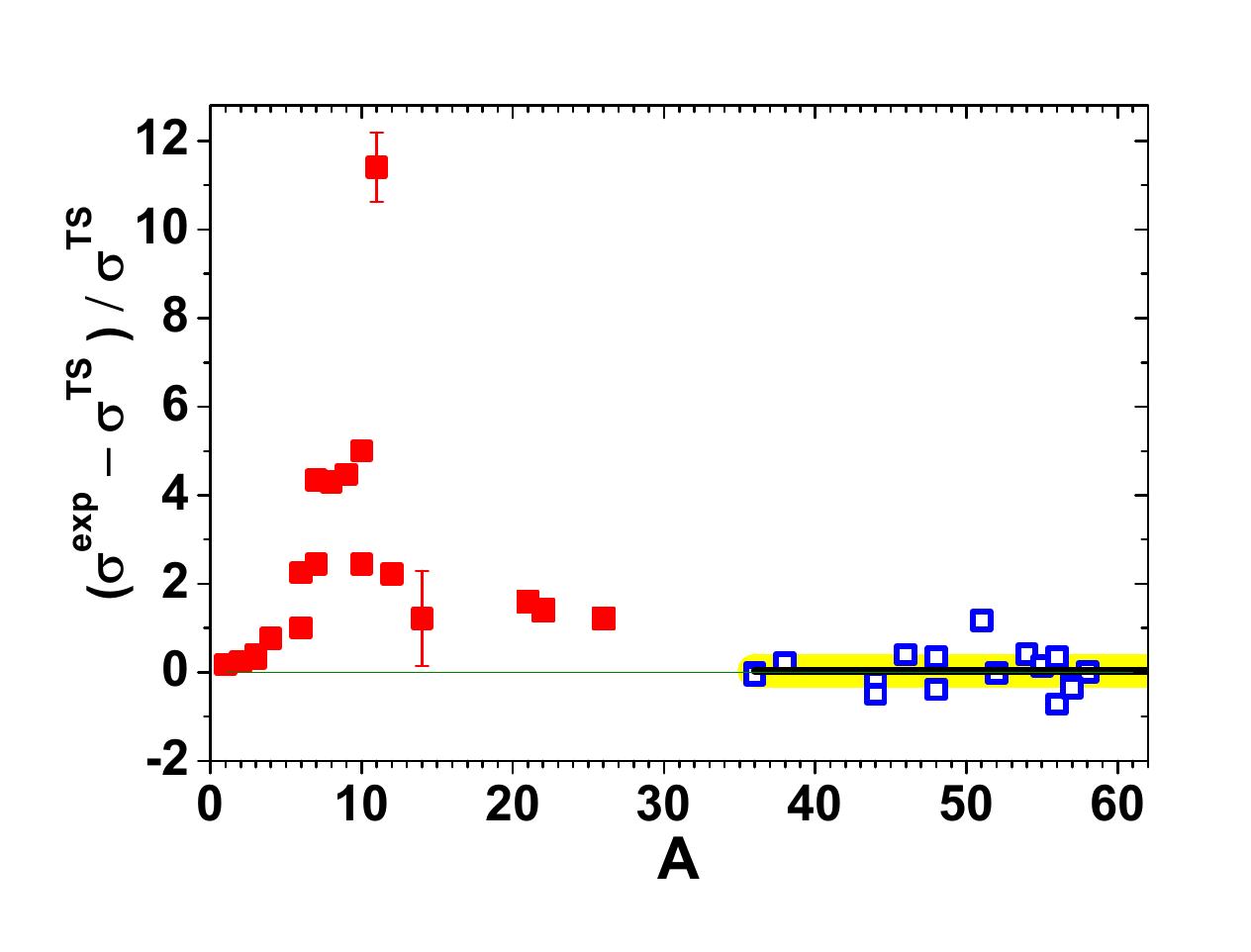}
\caption{\label{fig:relexpincl} (Color online) Relative difference
of experimental total production cross sections $\sigma^{exp}$ and
those from the two-step model $\sigma^{TS}$, averaged over beam
energy of 1 -- 3 GeV as a function of product mass.  The full
squares show total cross sections for products lighter than
$^{36}$Cl and  the open squares correspond to heavier products
\protect\cite{BUD09A,BUD09B,AMM08A,GRE88A,MIC95A,MIC84A,SCH96A,RAI75A,REG79A}.
 The horizontal
solid line drawn for those latter products shows the average value
of the relative difference of the cross sections, and the shadowed,
horizontal bar presents the standard deviation of the relative
difference.}
\end{center}
\end{figure}

To inspect more closely, for which products the break up
contribution is very significant, the differences of experimental
production cross sections and those evaluated by the two-step model
were for each product averaged over proton beam energies 1 -- 3 GeV,
and normalized to the energy averaged cross section of the two-step
model.  These relative differences are presented in Fig.
\ref{fig:relexpincl} as a function of the product mass. It is
clearly seen, that the relative differences are rather small and
\emph{of a random character} for products with mass larger than
A~$\sim$ 30.  Average value of the relative difference for these
products, shown as a horizontal, solid line, is very close to zero
(0.03)  and the standard deviation, depicted as a horizontal,
shadowed bar  is equal to 0.44. This points out that the two-step
model is able to reproduce quantitatively the total production cross
sections for heavy reaction products.
In contrast, a different  mass dependence of the relative difference
is observed for products lighter than A $\sim$ 30. In that mass
region the experimental cross sections are larger than those
calculated by two-step model \emph{for all products}, giving room
for the additional process. This indicates that for LCPs as well as
for IMFs with a mass smaller than A $\sim$ 30 the break-up
contribution sets in. This contribution is dominating for IMFs
because the relative difference is quite large for them; $\sim$~4 in
the neighborhood of mass A=10 being about a factor two smaller for
larger masses.  For hydrogen and helium isotopes, the relative
difference ($\sim$ 0.3) is smaller than for IMFs, thus the relative
break-up contribution is significant but not so large as for IMFs.

\section{Summary and conclusions}

Due to the recent determination of total production cross sections
for LCPs and light IMFs  at several proton beam energies (0.175,
1.2, 1.9, and 2.5 GeV) \cite{BUD09A,BUD09B} it became possible to
perform a systematic survey of the energy dependence of these cross
sections in the full range of product masses.  The excitation
functions are smooth and vary systematically with the mass of the
products.  The comparison of theoretical excitation functions,
evaluated in the frame of two-step model -- intranuclear cascade of
nucleon-nucleon collisions followed by an evaporation of particles
-- with the experimental data indicated, that the production of the
target-like particles (i.e., particles with A larger than $\sim$ 30)
is well reproduced by the two-step model. On the contrary, cross
sections predicted by this model for the production of lighter
nuclides are systematically several times smaller than the
experimental cross sections.

It was discussed, that the break-up of the target, proceeding in the
fast stage of the reaction, followed by an emission of particles
from the excited fragments of the target is able to explain the
observed mass dependence of the total production cross sections. The
same hypothesis was found to be very fruitful in the reproduction of
energy and angular dependencies of double differential cross
sections of LCPs and IMFs produced in p+Ni collisions at several
beam energies (0.175, 1.2, 1.9, and 2.5 GeV) \cite{BUD09A,BUD09B}.

  The presently existing models of the reaction are not able
to explain the observed effects, what was proved by extensive
studies of the production cross sections performed recently for the
p+$^{56}$Fe nuclear system, very similar to the p+Ni system studied
in the present work,
 by two groups: \cite{VIL07A} and \cite{TIT08A}.
In those papers the measured data have been compared with
calculation results of 15 different codes for hadron-nucleus
interactions: MCNPX (INCL, CEM2K, BERTINI, ISABEL), LAHET (BERTINI,
ISABEL), CEM03 (.01, .G1, .S1), LAQGSM03 (.01, .G1, .S1),
CASCADE-2004, LAHETO, and BRIEFF and still the agreement was not
satisfactory.  Moreover, the authors of Ref. \cite{TIT08A} claimed,
that "\emph{the most significant calculation-to-experiment
differences are observed in the yields of the A $<$ 30 light nuclei,
indicating that further improvements in nuclear reaction models are
needed}".

The present hypothesis of the break-up of the target nucleus, which
in a natural way explains the observed effects, offers an important
hint in the development of microscopic transport codes, which up to
now do not take such a process into account.

\begin{acknowledgments}


The technical support of A.Heczko, W. Migda{\l}, and N. Paul in
preparation of experimental apparatus is greatly appreciated.  This
work was supported by the European Commission through European
Community-Research Infrastructure Activity under FP6 project Hadron
Physics, contract number RII3-CT-2004-506078 as well as the FP6
IP-EUROTRANS FI6W-CT-2004-516520. This work was also partially
supported by the Helmholtz Association through funds provided to the
virtual institute "Spin and strong QCD" (VH-VI-231). One of us (MF)
acknowledges gratefully financial support of Polish Ministry of
Science and Higher Education (Grant No N N202 174735, contract
number 1747/B/H03/2008/35).

\end{acknowledgments}

\end{document}